\documentstyle[12pt,fleqn]{article}
\topmargin - 0.8cm

\catcode`@=11
\@addtoreset{equation}{section}
\catcode`@=12
\mathchardef\SGamma="7100

\begin{document}
\title{\bf Open inflation without anthropic principle}
\author{A.O. Barvinsky$^{\dag}$}
\date{}
\maketitle
\hspace{-8mm}{\em
Theory Department, Lebedev Physics Institute and Lebedev Research Center in
Physics, Leninsky Prospect 53,
Moscow 117924, Russia}
\begin{abstract}
We propose the mechanism of quantum creation of the open Universe in the
observable range of values of $\Omega$. This mechanism is based on the
no-boundary quantum state with the Hawking-Turok instanton in the model
with nonminimally coupled inflaton field and does not use any anthropic
considerations. Rather, the probability distribution peak with necessary
parameters of the inflation stage is generated on this instanton due to
quantum loop effects. In contrast with a similar mechanism for closed models,
existing only for the tunneling quantum state of the Universe, open
inflation originates from the no-boundary cosmological wavefunction.
\end{abstract}
$^{\dag}$e-mail: barvin@td.lpi.ac.ru\\

\section{Introduction}
\hspace{\parindent}
Hawking and Turok have recently suggested the mechanism of quantum
creation of an open Universe from the no-boundary cosmological state
\cite{HawkTur}. Motivated by the observational evidence for inflationary
models with $\Omega<1$ they constructed a singular gravitational
instanton capable of generating expanding universes with open spatially
homogeneous sections.
The prior quantum probability of such universes weighted by the anthropic
probability of galaxy formation was shown to be peaked at $\Omega\sim 0.01$.
This idea, despite its extremely attractive nature, was criticized from
various sides. In order to increase the amount of inflation to larger values
of $\Omega$ and avoid anthropic considerations Linde \cite{Lindop} proposed
to replace the no-boundary quantum state \cite{HH,H} by the tunneling one
\cite{VilNB,tun}. The singularity
of the Hawking-Turok instanton raised a number of objections both in the
Euclidean theory \cite{BousLind,Vilen,Unruh,Wu,Garriga} and from the
viewpoint of the resulting timelike singularity in the expanding Universe
\cite{Unruh,Starobin,Garriga}. The criticism of singular instantons was
followed by attempts of their justification \cite{HawkTur1,Garriga1} which
still leave their issue open.

In any case it seems that the practical goal of quantum cosmology --
generating the open Universe with observationally justified modern value of
$\Omega$, not very close to one or zero, -- has not yet been reached. The use
of anthropic principle, as was recognized by the authors of \cite{HawkTur},
is certainly a retreat in theory, because by and large this principle has
such a disadvantage that it can explain practically everything without
being able to predict anything. The tunneling state advocated by Linde
\cite{Lindop} (and strongly criticized in \cite{HawkTur2}) requires special
supergravity induced potentials and takes place at energies beyond reliable
perturbative domain with the resulting $\Omega\simeq 1$. Other works in the
above series discuss conceptual issues of the Hawking-Turok proposal
without offering the concrete mechanism of generating the needed $\Omega$.

On the other hand, in spatially {\it closed} context there exists a
mechanism of
generating the probability peak in the cosmological wavefunction at a low
(typically GUT) energy scale. It does not appeal to anthropic considerations.
Rather it is based on quantum loop effects \cite{norm,tunnel} in the
model of chaotic inflation with large negative nonminimal coupling of the
inflaton \cite{qsi,qcr,efeq}. In the quantum gravitational domain the
conventional expression for the no-boundary and tunneling probability
distributions of the inflaton field
$\rho_{\rm NB,T}(\varphi)\sim\exp[\mp I(\varphi)]$ is replaced by
	\begin{eqnarray}
	\rho_{\rm NB,T}(\varphi)\sim\exp[\mp I(\varphi)-
	\mbox{\boldmath$\SGamma$}(\varphi)],          \label{0.1}
	\end{eqnarray}
where the classical Euclidean action $I(\varphi)$ on the quasi-DeSitter
instanton with
the inflaton value $\varphi$ is amended by the loop effective action
$\mbox{\boldmath$\SGamma$}(\varphi)$ calculated on the same instanton
\cite{norm,tunnel}. The contribution of the latter can qualitatively change
predictions of the tree-level theory due to the dominant anomalous scaling
part of the effective action. On the instanton of the size $1/H(\varphi)$ --
the inverse of the Hubble constant, it looks like
$\mbox{\boldmath$\SGamma$}(\varphi)\sim Z\ln H(\varphi)$ where $Z$ is the
total anomalous scaling of all quantum fields in the model. For the model
of \cite{qsi,qcr}
	\begin{equation}
	{\mbox{\boldmath $L$}}(g_{\mu\nu},\varphi)
	=g^{1/2}\left\{\frac{m_{P}^{2}}{16\pi} R(g_{\mu\nu})
	-\frac{1}{2}\xi\varphi^{2}R(g_{\mu\nu})
	-\frac{1}{2}(\nabla\varphi)^{2}
	-\frac{1}{2}m^{2}\varphi^{2}
	-\frac{\lambda}{4}\varphi^{4}\right\},      \label{0.2}
	\end{equation}
with a big negative constant $-\xi=|\xi|\gg 1$ of nonminimal curvature
coupling, and generic GUT sector of Higgs $\chi$, vector gauge $A_\mu$
and spinor fields $\psi$ coupled to the inflaton via the interaction term
	\begin{eqnarray}
	{\mbox{\boldmath $L$}}_{\rm int}
	=\sum_{\chi}\frac{\lambda_{\chi}}4
	\chi^2\varphi^2
	+\sum_{A}\frac12 g_{A}^2A_{\mu}^2\varphi^2+
        \sum_{\psi}f_{\psi}\varphi\bar\psi\psi
	+{\rm derivative\,\,coupling},             \label{0.3}
	\end{eqnarray}
this parameter can be very big, because it is quadratic in $|\xi|$,
$Z=6|\xi|^2\mbox{\boldmath$A$}/\lambda$ with a universal combination of
the coupling constants above
	\begin{eqnarray}
	{\mbox{\boldmath $A$}} = \frac{1}{2\lambda}
	\Big(\sum_{\chi} \lambda_{\chi}^{2}
	+ 16 \sum_{A} g_{A}^{4} - 16
	\sum_{\psi} f_{\psi}^{4}\Big).               \label{A}
	\end{eqnarray}

Thus, the probability peak in this model reduces to the extremum of
the function
	\begin{eqnarray}
	\ln\rho_{\rm NB,\,T}(\varphi)\simeq\mp I(\varphi)
	-3\frac{|\xi|^2}\lambda\mbox{\boldmath$A$}\,
	\ln\frac{\varphi^2}{\mu^2}.                    \label{1.1}
	\end{eqnarray}
in which the $\varphi$-dependent part of the classical instanton action
	\begin{eqnarray}
	&&I(\varphi)=-\frac{96\pi^2|\xi|^2}\lambda
	-\frac{24\pi(1+\delta)|\xi|}
	{\lambda}\frac{m_P^2}{\varphi_0^2}
	+O\,\left(\frac{m_P^4}{\varphi^4}\right),     \label{1.2} \\
	&&\delta\equiv
	-\frac{8\pi\,|\xi|\,m^2}{\lambda\,m_P^2},    \label{delta}
	\end{eqnarray}
should be balanced by the anomalous scaling term provided the signs
of $(1+\delta)$ and $\mbox{\boldmath$A$}$ are properly correlated with
the $(\mp)$ signs of the no-boundary (tunneling) proposals. As a result
the probability peak exists with parameters -- mean values of the inflaton
and Hubble constants and relative width
	\begin{eqnarray}
	&&\varphi_I^2= m_{P}^2\frac{8\pi|1+\delta|}{|\xi|
	{\mbox{\boldmath$A$}}},\,\,\,\,\,
	H^2(\varphi_I)=
        m_{P}^2\frac{\lambda}{|\xi|^2}
	\frac{2\pi|1+\delta|}
	{3{\mbox{\boldmath $A$}}},                  \label{1.3} \\
	&&\frac{\Delta\varphi}{\varphi_I}\sim
	\frac{\Delta H}{H}\sim
	\frac 1{\sqrt{12{\mbox{\boldmath $A$}}}}
        \frac{\sqrt{\lambda}}{|\xi|},              \label{1.4}
	\end{eqnarray}
which are strongly suppresed by a small ratio $\sqrt{\lambda}/|\xi|$
known from the COBE normalization for $\Delta T/T\sim 10^{-5}$
\cite{COBE,Relict}(because
the CMBR anisotropy in this model is proportional to this ratio
\cite{nonmin}). This GUT scale peak gives rise to the finite
inflationary epoch with the e-folding number
	\begin{eqnarray}
	N\simeq \frac{48\pi^2}{\mbox{\boldmath $A$}}.   \label{1.5}
	\end{eqnarray}
only for $1+\delta>0$ and, therefore, only for the {\it tunneling} quantum
state (plus sign in (\ref{1.1})). Comparison with $N\geq 60$ necessary
for $\Omega>1$ immeadiately yields the bound on
$\mbox{\boldmath $A$}\sim 5.5$ \cite{efeq} which can be regarded as a
selection criterion for particle physics models \cite{qsi}. This conclusions
on the nature of the inflation dynamics from the initial probability peak
remain true also at the quantum level -- with the effective equations
replacing the classical equations of motion \cite{efeq}.

For the proponents of the no-boundary vs tunneling quantum states this
situation might seem unacceptable. According to this result the
no-boundary proposal does not generate realistic inflationary scenario,
while the tunneling state does not satisfy important aesthetic criterion
-- the universal formulation of both the initial conditions and dynamical
aspects in one concept -- spacetime covariant path integral over geometries.
The criticism of the tunneling state in \cite{HawkTur2} is not completely
justified, because as a solution of the Wheeler-DeWitt equation this
state can be constructed as a normalizable (gaussian) vacuum of linearized
inhomogeneous modes, see \cite{VilVach,tvsnb}. But this construction,
apparently, cannot be achieved by a sort of Wick rotation in the spacetime
covariant path integral without breaking important locality properties
\cite{HawkTur2}.

In this paper we show that for the {\it open} Universe the situation
qualitatively reverses: the probability peak of the quantum (one-loop)
distribution of the open inflationary models exists for the no-boundary
state based on the Hawking-Turok instanton. Similarly to (\ref{1.3}) -
(\ref{1.5}) it has GUT scale parameters and the value of $N$ easily
adjustable (without fine tuning of initial conditions and anthropic
considerations) for observationally justified values of $\Omega$. For this
purpose in the next section we develop the slow-roll approximation
technique for the Hawking-Turok instanton with a minimal inflaton. In Sect.3
we extend this technique to the nonminimally coupled inflaton and obtain
the instanton action to the subleading order in the slow-roll
parameter. Remarkably, this tree-level action features large positive
contribution logarithmic in the inflaton field structurally analogous to
loop corrections. In Sect.4 we discuss the difficulties with the quantum
effective action due to the singularity of the Hawking-Turok instanton and
establish that the dominant scaling behaviour is robust against this
singularity. Finally, in Sect.5 we calculate the most probable $H$, $N$
and $\Omega$ of the open inflation generated within the no-boundary
Hawking-Turok paradigm.

\section{Minimal inflaton}
\hspace{\parindent}
We assume that the reader is familiar with the construction of the
Hawking-Turok instanton in a simple model with the minimally coupled
inflaton field \cite{HawkTur}
	\begin{eqnarray}
	I[g_{\mu\nu},\varphi]=\int_{\cal M} d^4x\, g^{1/2}\left\{
	-\frac{m_P^2}{16\pi}R+\frac12(\nabla\varphi)^2+V(\varphi)\right\}
	+\frac{m_P^2}{8\pi}
	\int_{\partial\cal M}d^3x\, ({}^3g)^{1/2} K.    \label{2.1}
	\end{eqnarray}
Here $V(\varphi)$ is the inflaton potential, $K$ is the extrinsic curvature
of the spacetime boundary at the point of the singularity in the
Euclidean spacetime domain. With the spatially homogeneous ansatz for the
Euclidean metric ($d\Omega^2_{(3)}$ is the metric of the 3-dimensional sphere
of unit radius and $a(\sigma)$ is the scale factor),
	\begin{eqnarray}
	ds^2=d\sigma^2+a^2(\sigma)\,d\Omega^2_{(3)} ,    \label{2.2}
	\end{eqnarray}
the Euclidean equations of motion take the form
	\begin{eqnarray}
	&&\ddot\varphi+3\frac{\dot a}a\dot\varphi-V'=0,  \label{2.3}\\
	&&a^3V-\frac{3m_P^2}{8\pi}(a-a\dot a^2)
	-\frac12 a^3\dot\varphi^2=0,                      \label{2.4}
	\end{eqnarray}
where dots denote the derivatives with respect to the coordinate $\sigma$
and the prime denotes the derivative with respect to the inflaton scalar
field. In the vicinity of the point $\sigma=0$ where $\dot\varphi(0)=0$ and
$a(\sigma)\sim \sigma$ the solution of these equations can be obtained
by the slow-roll expansion in powers of the gradient of the inflaton
potential. In the lowest order approximation this is the constant
inflaton field $\varphi_0=\varphi(0)$ and the Euclidean DeSitter geometry
	\begin{eqnarray}
	&&a=\frac1{H_0} \sin\theta+\delta a,\,\,\,
	\varphi=\varphi_0+\delta\varphi,                \label{2.5} \\
	&&H_0^2=\frac{8\pi V(\varphi_0)}{3m_P^2},\,\,\,
	\theta=H_0\sigma,                            \label{2.6}
	\end{eqnarray}
with the effective Hubble constant $H_0$ given in terms of the constant
potential $V(\varphi_0)$.

When the slope of the inflaton potential is not too steep one can apply
the slow roll expansion in powers of the parameter
	\begin{eqnarray}
	\varepsilon=\frac1{\sqrt{3\pi}}
	\frac{V'(\varphi_0)}{V(\varphi_0)}                     \label{2.7}
	\end{eqnarray}
($|\dot V/HV|\simeq 3\epsilon^2/8$) to find the first-order approximation
(see \cite{qcr} for details)
	\begin{eqnarray}
	&&\delta\varphi(\theta)=\sqrt{\frac3{16\pi}}m_P
	\varepsilon
	\left(\frac14 \tan^2\frac\theta2
	-\ln\cos\frac\theta2\right),              \label{2.8}\\
	&&\delta a(\theta)=\frac1{H_0}\,
	O\left(\varepsilon^2\right).                  \label{2.9}
	\end{eqnarray}

As shown in \cite{HawkTur}, for monotonically growing potentials the scale
factor of the solution, starting at $\sigma=0$ with the initial conditions
of the above type, vanishes at some $\sigma=\sigma_f$, $a(\sigma_f)=0$,
and the behaviour of fields near this point have the form
	\begin{eqnarray}
	a\simeq A(\sigma_f-\sigma)^{1/3},\,\,\,
	\varphi\simeq-\frac{m_P}{\sqrt{12\pi}}
	\ln(\sigma_f-\sigma)+\Phi_0,\,\,\,
	\sigma\rightarrow\sigma_f.                 \label{2.10}
	\end{eqnarray}
It is important that the coefficient of the logarithmic singularity of the
scalar field is unambiguously defined from the equations of motion, whereas
the coefficients $A$ and $\Phi_0$ nontrivially depend on the initial condition
at $\sigma=0$, that is on $\varphi_0$. To find them as functions of
$\varphi_0$ we develop the perturbation expansion of the solution near
$\sigma_f$. In contrast with the slow roll expansion near $\sigma=0$ this is
the expansion in powers of the potential $V(\varphi)$ itself rather than
its gradient, and $\dot\varphi$-derivatives give the dominant contribution
at this asymptotics. Then we match the both asymptotic expansions in the
domain of $\sigma$ where they are both valid (it turns out that such a domain
really exists and corresponds to the range of the angular coordinate $\theta$
in eqs.(\ref{2.5})-(\ref{2.9}), $1\gg \pi-\theta\gg\varepsilon^{1/2}$, where
the corrections (\ref{2.8})-(\ref{2.9}) are small). From this match one easily
finds all the unknown parameters $A,\Phi_0,\sigma_f$ as functions of
$\varphi_0$. Omitting the details which will be published elsewhere we give
here the result in the lowest order of the slow roll expansion
	\begin{eqnarray}
	&&\theta_f\equiv H_0\sigma_f\simeq\pi
	-\frac{2\pi^{3/2}}{\Gamma^2(1/4)}
	\varepsilon^{1/2},                          \label{2.11}\\
	&&A\simeq\left(\frac{3\varepsilon}
	{H_0^2}\right)^{1/3},\\
	&&\Phi_0\simeq\varphi_0-\frac12\frac{m_P}{\sqrt{12\pi}}
	\ln\left[\frac{9H_0^2}{8\varepsilon}\right]
	=\varphi_0+\frac12\frac{m_P}{\sqrt{12\pi}}
	\ln\frac{V_0'}{V_0^2}+{\rm const}.                \label{2.12}
	\end{eqnarray}

The knowledge of $A(\varphi_0$ and $\Phi_0(\varphi_0)$ allows one to
obtain the action of the Hawking-Turok instanton. Its classical Euclidean
action
	\begin{eqnarray}
	I(\varphi_0)=\left.2\pi^2\int_0^{\sigma_f}d\sigma\,\left\{a^3V
	-\frac{3m_P^2}{8\pi}a(1+\dot a^2)
	+\frac12 a^3\dot\varphi^2\right\}
	\right|_{\varphi(\sigma,\varphi_0),\,\,
	a(\sigma,\varphi_0)}                           \label{2.13}
	\end{eqnarray}
depends on $\varphi_0$ and the mechanism of this dependence
originates from the behaviour of fields at the singularity. Indeed,
differentiating (\ref{2.13}) with respect to
$\varphi_0$ one finds that the volume part vanishes in virtue of equations
of motion, while integrations by parts give a typical surface term involving
the Lagrangian of (\ref{2.13}) and its derivatives with respect to
$(\dot\varphi,\dot a)$, which does not vanish at $\sigma_f$.

For $\sigma$ close to $\sigma_f$ the functional dependence of variables
(\ref{2.10}) on $\varphi_0$ enters through the coefficients $A,\,\Phi_0$ as
well as through $\sigma_f$. Since $\sigma_f$ enters the fields in the
combination $\sigma-\sigma_f$, the total derivative of the field takes
the form $d\varphi/d\varphi_0=\partial\varphi/\partial\varphi_0
-\dot\varphi(\partial\sigma_f/\partial\varphi_0)$, where partial derivative
with respect to $\varphi_0$ acts only on $\Phi_0$ (and coefficients of higher
powers in $(\sigma-\sigma_f)$). Similar relation holds also for the scale
factor. Thus, the surface term at $\sigma_f$ equals
	\begin{eqnarray}
	&&\frac{dI(\varphi_0)}{d\varphi_0}=\left.\left(L
	-\dot\varphi\frac{\partial L}{\partial\dot\varphi}
	-\dot a\frac{\partial L}{\partial\dot a}\right)
	\frac{\partial\sigma_f}{\partial\varphi_0}\,
	\right|_{\,\sigma_f}\nonumber\\
	&&\qquad\qquad\qquad
	+\left[\,\frac{\partial L}{\partial\dot a}
	\frac{\partial A}{\partial\varphi_0}
	(\sigma_f-\sigma)^{1/3}
	+\frac{\partial L}{\partial\dot\varphi}
	\frac{\partial\Phi_0}{\partial\varphi_0}\right]
	_{\,\sigma\rightarrow\sigma_f},                  \label{2.14}
	\end{eqnarray}
where $L$ is the Lagrangian of the Euclidean action (\ref{2.13}). The first
term here identically vanishes, because it is proportional to the Hamiltonian
constraint (in terms of velocities). On using (\ref{2.10}) - (\ref{2.12})
one then finds
	\begin{eqnarray}
	\frac{dI(\varphi_0)}{d\varphi_0}\simeq\left(\frac d{d\varphi_0}+
	\sqrt{\frac3{16\pi}}m_P\frac{d^2}
	{d\varphi_0^2}\right)\,
	\left(-\frac{3m_P^4}{8V(\varphi_0)}\right),      \label{2.15}
	\end{eqnarray}
whence in the first order of the slow roll expansion
	\begin{eqnarray}
	I(\varphi_0)\simeq\left(1+\sqrt{\frac3{16\pi}}m_P
	\frac d{d\varphi_0}\right)\,
	\left(-\frac{3m_P^4}{8V(\varphi_0)}\right).    \label{2.16}
	\end{eqnarray}
This expression reproduces the result of ref.\cite{HawkTur1} obtained by
indirect and less rigorous method\footnote
{The calculations of \cite{HawkTur1} do not take into account the slow roll
corrections to the volume part of the action.}.
One can check that the second term corresponds to the
contribution of the extrinsic curvature surface part of the action
(\ref{2.1}). A remarkable property of the obtained algorithm is that it
is universal for a wide class of inflaton potentials $V(\varphi_0)$
($V(\varphi)$ should only satisfy typical restrictions imposed by the
slow roll expansipon) and in a closed form expresses the result in terms
of $V(\varphi)$ and its derivative.

\section{Nonminimal coupling}
\hspace{\parindent}
We shall be interested in the action with the nonminimal inflaton field
coupled to curvature via the $\varphi$-dependent Planck ``mass''
$16\pi U(\varphi)$
	\begin{eqnarray}
	I=\int_{\cal M} d^4x\, g^{1/2}\left\{V(\varphi)
	-U(\varphi)R+\frac12(\nabla\varphi)^2\right\}
	+2\int_{\partial\cal M}d^3x\,
	({}^3g)^{1/2} U(\varphi)K.                    \label{3.1}
	\end{eqnarray}
It is well known that this action can be reparametrized to the Einstein frame
by special conformal transformation and reparametrization of the
inflaton field $(g_{\mu\nu},\varphi)\rightarrow(G_{\mu\nu},\phi)$. These
transformations are implicitly given by equations \cite{renorm}
	\begin{eqnarray}
	&&G_{\mu\nu}=\frac{16\pi U(\varphi)}
	{m_P^2}g_{\mu\nu},                            \label{3.2}\\
	&&\left(\frac{d\phi}{d\varphi}\right)^2
	=\frac{m_P^2}{16\pi}\frac{U+3U'^2}{U^2}.      \label{3.3}
	\end{eqnarray}
The action in terms of new fields
	\begin{eqnarray}
	\bar{I}=\int_{\cal M} d^4x\, G^{1/2}\left\{\bar{V}(\phi)
	-\frac{m_P^2}{16\pi}R(G_{\mu\nu})
	+\frac12(\bar{\nabla}\phi)^2\right\}
	+\frac{m_P^2}{8\pi}\int_{\partial\cal M}
	d^3x\, ({}^3G)^{1/2} \bar{K}                  \label{3.4}
	\end{eqnarray}
has a minimal coupling and the new inflaton potential
	\begin{eqnarray}
	\bar{V}(\phi)=\left.\left(\frac{m_P^2}{16\pi}\right)^2
	\frac{V(\varphi)}{U^2(\varphi)}
	\,\right|_{\varphi=\varphi(\phi)}.            \label{3.5}
	\end{eqnarray}
The bar indicates here that the corresponding quantity is calculated in the
Einstein frame of fields $(G_{\mu\nu},\phi)$.

The above transition to the Einstein frame allows one to find the
Hawking-Turok instanton for the model (\ref{3.1}) by transforming the
results of the previous section to the nonminimal frame. Here we shall do it
in the case of a big negative nonminimal coupling $|\xi|\gg 1$:
	\begin{eqnarray}
	U(\varphi)=\frac{m_P^2}{16\pi}
	+\frac12 |\xi|\,\varphi^2                    \label{3.6}
	\end{eqnarray}
and quartic potential of (\ref{0.2}). The integration of eq.(\ref{3.3}) for
large values of the inflaton field, $|\xi|\,\varphi^2/m_P^2\gg 1$, expresses
$\varphi$ in terms of Eistein frame field $\phi$
	\begin{eqnarray}
	\varphi(\phi)\simeq\frac{m_P}{|\xi|^{1/2}}
	\exp\left[\sqrt{4\pi/3}
	\Big(1+\frac1{6\,|\xi|}\Big)^{-1/2}
	\frac\phi{m_P}\right],                       \label{3.7}
	\end{eqnarray}
where the integration constant is chosen so that the above range of $\varphi$
corresponds to $\phi\gg m_P$. The potential in the Einstein frame
equals
	\begin{eqnarray}
	\bar{V}(\phi)=\frac{\lambda m_P^4}{256\pi^2|\xi|^2}
	\,\left[\,1-\frac{1+\delta}{4\pi}\frac{m_P^2}
	{|\xi|\,\varphi^2}+...\right]
	_{\,\varphi=\varphi(\phi)},                  \label{3.8}
	\end{eqnarray}
where we have retained only the first order term in $m_P^2/|\xi|\,\varphi^2$.
In view of (\ref{3.7}), for large $\phi$ this potential exponentially
approaches a constant and satisfies a slow roll approximation with the
expansion parameter
	\begin{eqnarray}
	\varepsilon=\frac{m_P}{\sqrt{3\pi}}\frac{\bar V'(\phi_0)}
	{\bar V(\phi_0)}
	\simeq\frac{1+\delta}{3\pi}
	\left(1+\frac1{6\,|\xi|}\right)^{-1/2}
	\frac{m_P^2}{|\xi|\,\varphi_0^2}\ll 1.          \label{3.9}
	\end{eqnarray}

This justifies the above choice of range for the values of the inflaton field.
In this range the Hawking-Turok instanton is described by the equations of the
previous section for the Einstein frame fields $\bar a(\bar\sigma)$ and
inflaton $\phi(\bar\sigma)$. Here $\bar\sigma$ is the coordinate in the
spacetime interval $d\bar s^2=d\bar\sigma^2+\bar a^2(\bar\sigma)\,
d\Omega^2_{(3)}$ of the Einstein frame metric. In view of (\ref{3.2}) these
intervals are related by the equation $d\bar s^2=(16\pi U/m_P^2) ds^2$, so
that the coordinates and scale factors of both frames\footnote
{Note that the relation (\ref{3.2}) holds in one coordinate system covering
the both conformally related spacetimes, while the coordinates $\bar\sigma$
and $\sigma$ are essentially different.}
are related by $d\bar\sigma=\sqrt{16\pi U/m_P^2}\,d\sigma$ and
$\bar a=\sqrt{16\pi U/m_P^2}\,a$. Combining these equations with the asymptotic
behaviour of the Einstein frame fields at $\bar\sigma\rightarrow\bar\sigma_f$
(eqs. (\ref{2.10})-(\ref{2.12}) rewritten for
$\bar a(\bar\sigma),\phi(\bar\sigma)$ with the potential $\bar{V}(\phi)$)
one can easily find the behaviour of
fields in the nonminimal frame. We give it in the limit of large $|\xi|$:
	\begin{eqnarray}
	&&a(\sigma)\simeq\frac4{m_P}
	\left(\frac{m_P}{\varphi_0}\right)^{1+2\epsilon}
	\left(\frac{1+\delta}{4\pi\lambda}\right)
	^{1/4+\epsilon/2}\,
	\Big[\,m_P(\sigma_f-\sigma)\,\Big]^{1/2-\epsilon}, \label{3.10}\\
	&&\varphi(\sigma)\simeq m_P
	\left(\frac{\varphi_0}{m_P}\right)^{1/2+3\epsilon}
	\left(\frac{1+\delta}{4\pi\lambda}\right)^{1/8-3\epsilon/4}\,
	\Big[\,m_P(\sigma_f-\sigma)\,\Big]^{-1/4+3\epsilon/2},\\
	&&\epsilon\equiv\frac12\,\frac{\sqrt{1+1/6\,|\xi|}-1}
	{1+3\sqrt{1+1/6\,|\xi|}}\simeq\frac1{96\,|\xi|}\ll 1. \label{3.11}
	\end{eqnarray}
In contrast with the minimal coupling we now have the power singularities
for both fields. For large $|\xi|\gg 1$, in particular, they look like
$a\sim (\sigma_f-\sigma)^{1/2}$ and $\varphi\sim (\sigma_f-\sigma)^{-1/4}$.
The inflaton singularity is thus stronger than the logarithmic one in the
minimal case, while that of the scale factor is softer
($1/2-\epsilon\geq 1/3$). Note, by the way, that the coefficient of
strongest singularity of the scalar curvature is also suppressed by $1/|\xi$,
$R\sim (1/|\xi|)(\sigma-\sigma_f)^{-2}$. This
property can be qualitatively explained by the fact that the effective
gravitational constant $(m_P^2+8\pi|\xi|\varphi^2)^{-1}$ tends to zero at
the singularity.

The classical action can also be easily calculated in the Einstein frame
$I(\varphi_0)=\bar I(\phi_0)$ with the aid of eq.(\ref{2.16}). Taking into
account that
	\begin{eqnarray}
	\frac{3m_P^4}{8\bar V(\phi_0)}\simeq\frac{96\pi^2|\xi|^2}\lambda
	+\frac{24\pi(1+\delta)|\xi|}{\lambda}\frac{m_P^2}{\varphi_0^2}
	+\frac32\frac{(1+2\delta)^2}\lambda
	\left(\frac{m_P^2}{\varphi_0^2}\right)^2+...        \label{3.12}
	\end{eqnarray}
and using the relation
	\begin{eqnarray}
	\sqrt{\frac3{16\pi}}m_P\frac d{d\phi}\simeq\frac12
	\left(1+\frac1{6\,|\xi|}\right)^{-1/2}\varphi\frac d{d\varphi},
	\end{eqnarray}
one finds that the surface term at the singularity (the term with the
derivative in (\ref{2.16})) almost cancels the first subleading
term in the expansion (\ref{3.12}) and inverts the sign of the second order
term
	\begin{eqnarray}
	&&\left(1+\sqrt{\frac3{16\pi}}m_P
	\frac d{d\phi_0}\right)\,
	\left(-\frac{3m_P^4}{8\bar{V}(\phi_0)}\right)\simeq
	-\frac{96\pi^2|\xi|^2}\lambda
	-\frac{2\pi(1+\delta)}\lambda \frac{m_P^2}{\varphi_0^2}\nonumber\\
	&&\qquad\qquad\qquad\qquad+\frac32\frac{(1+2\delta)^2}\lambda
	\left(\frac{m_P^2}{\varphi_0^2}\right)^2+...\,.    \label{3.13}
	\end{eqnarray}
Thus, in contrast with the closed model, these terms (of different
powers in $m_P^2/\varphi^2$) are of the same order of magnitude in $1/|\xi|$.
Later we shall see that at the probability maximum $m_P^2/\varphi^2\gg 1$
(although $m_P^2/|\xi|\varphi^2\sim\varepsilon\ll 1$) which means that
the dominant effect comes from the third term of (\ref{3.13}). This term is
however not reliable unless we take into account the complete subleading
order of the slow roll expansion in $\varepsilon$. Obtaining it is rather
cumbersome and is not so universal as in the lowest order, because the
result depends on a particular form of the inflaton potential. Without
going into details which will be published elsewhere we present the result
for our model:
	\begin{eqnarray}
	&&\theta_f\equiv H_0\sigma_f\simeq\pi
	-\frac{2\pi^{3/2}}{\Gamma^2(1/4)}
	\varepsilon^{1/2}-\frac{5\Gamma^2(1/4)}{48\pi^{1/2}}
	\varepsilon^{3/2},                                \label{3.14}\\
	&&A^3\simeq\frac{3\varepsilon}{H_0^2}
	-\frac{11}4\frac{\varepsilon^2}{H_0^2},\\
	&&\Phi_0\simeq\varphi_0-\frac12\frac{m_P}{\sqrt{12\pi}}
	\ln\left[\frac{9H_0^2}{8\varepsilon}\right]
	-\frac12\sqrt{\frac3{16\pi}}m_P\,\varepsilon
	\left(\ln\frac{\varepsilon}8
	+\frac{35}{18}\right),                       \label{3.15}
	\end{eqnarray}
where $\varepsilon=[(1+\delta)/3\pi]m_P^2/\varphi_0^2$. Using these
expressions for $A(\varphi_0)$ and $\Phi_0(\varphi_0)$ one obtains the
second order approximation for the Hawking-Turok action
	\begin{eqnarray}
	&&I_{\rm HT}(\varphi)=-\frac{96\pi^2|\xi|^2}\lambda
	-\frac{2\pi(1+\delta)}\lambda \frac{m_P^2}{\varphi_0^2}\nonumber\\
	&&\qquad+\frac{(1+\delta)^2}\lambda
	\left(\frac{m_P^2}{\varphi_0^2}\right)^2
	\left[\frac32\frac{(1+2\delta)^2}{(1+\delta)^2}-\frac{22}3
	+2\ln\left(\frac{24\pi|\xi|\varphi^2}
	{m_P^2(1+\delta)}\right)\right]
	+O\,\left(\frac{m_P^6}{|\xi|\varphi^6}\right).  \label{3.16}
	\end{eqnarray}
Due to big $|\xi|$ it contains a large but slowly varying (in $\varphi$)
logarithmic coefficient. The positive coefficient of this logarithmic
term actually follows from the sign of $\ln\cos(\theta/2)$ in the
equation (\ref{2.8}) for $\delta\varphi$ above and, thus, it is pretty well
fixed. This sign will have important consequences for quantum
creation of the open Universe. Note, by the way, that the logarithmic
structure of the result resembles the behaviour of Coleman-Weinberg
loop effective potentials (up to inversion of $\varphi$), even though this
term is entirely of a tree-level origin. Thus, the classical theory somehow
feels quantum structures when probing Planckian scales near the singularity.
This sounds rather coherent with recent results on holographic principle
in string theory when the tree-level theory in the bulk of spacetime
generates quantum theory on the boundary surface \cite{Witten}.

\section{Quantum corrections}
\hspace{\parindent}
The most vulnerable point of the Hawking-Turok instanton is the construction
of quantum corrections on its singular background. Although the classical
Euclidean action is finite, the quantum part of the effective action involving
the higher order curvature invariants is infinite, because their spacetime
integrals are not convergent at the singularity. At least
naively, this means that the whole amplitude is either suppressed to zero or
infinitely diverges indicating strong instability. Clearly, a self-consistent
treatment should regularize the arising infinities due
to the back reaction of the infinitely growing quantum stress tensor.

The result of such a self-consistent treatment is hardly predictable because
we do not yet have for it an exhaustive theoretical framework. This
framework might include fundamental stringy structures underlying our local
field theory, which are probed by Planckian curvatures near the singularity.
However, even without the knowledge of this fundamental framework it is worth
considering usual quantum corrections due to local fields
on a given singular background. This might help revealing those dominant
mechanisms that are robust against the presence of singularities and their
regulation due to back reaction and fundamental strings.

These quantum corrections can be devided into two main categories -- nonlocal
contributions due to massless or light degrees of freedom and local
contributions due to heavy massive fields \cite{beyond}. The effects from the first
category can be exactly calculable when they are due to the conformal anomaly
of the conformal invariant fields. For the Hawking-Turok instanton this
calculation can be based on the (singular) conformal transformation mapping
its geometry to the regular metric $d\tilde{s}^2$ of the half-tube
$R^+\times S^3$,
	\begin{eqnarray}
	ds^2=a^2(\sigma(X))\,d\tilde{s}^2,\,\,\,\,
	d\tilde{s}^2=dX^2+d\Omega_{(3)}^2,               \label{4.1}
	\end{eqnarray}
with the conformal coordinate $X=\int_\sigma^{\sigma_f}d\sigma'/a(\sigma'),\,
0\leq X<\infty$. With this conformal decomposition of the metric the effective
action $\SGamma[\,g_{\mu\nu}]$ can be represented as a sum of the finite
effective action of $\tilde{g}_{\mu\nu}$, $\SGamma[\,\tilde{g}_{\mu\nu}]$,
and $\Delta\SGamma[\,\tilde{g}_{\mu\nu},a]$ -- the anomalous action obtained
by integrating the known conformal anomaly along the orbit of the local
conformal group joining $g_{\mu\nu}$ and $\tilde{g}_{\mu\nu}$. The anomalous
action $\Delta\SGamma[\,\tilde{g}_{\mu\nu},a]$ is known for problems without
boundaries \cite{Riegert,Tsconf,BMZ}. For a singular conformal factor
$a^2(\sigma(X))$ the bulk part of this action is divergent, but the question
of its finiteness is still open, because in problems with
boundaries the conformal anomaly should have surface (simple and double
layer) contributions that might lead to finite anomalous action on the
Hawking-Turok instanton \cite{Bconf}.

Fortunately, the problem with large nonminimal coupling of the inflaton
falls into the second category of problems -- local effective action of
heavy massive fields. Due to the Higgs mechanism for all matter fields
interacting with inflaton by (\ref{0.3}) their particles acquire masses
$m^2\sim\varphi^2$ strongly exceeding the spacetime curvature
$R\sim\lambda\varphi^2/|\xi|$ \cite{qsi,qcr,efeq}. The renormalized
effective action expanded in powers of the curvature to mass squared ratio
$R/m^2\sim\lambda/|\xi|\ll 1$ for generic spacetime background has the
following form of the local Schwinger-DeWitt expansion \cite{DW,BarvV,efeq}
	\begin{eqnarray}
	&&\mbox{\boldmath$\SGamma$}^{\rm 1-loop}=
	-\frac1{32\pi^2}\,\int d^4x\,g^{1/2}\,
	{\rm tr}\,\left\{\frac12\left(\frac32-\ln\frac{m^2}{\mu^2}
	\right)\,m^4\hat{1}\right.\nonumber\\
	&&\quad\qquad+\left(1-\ln\frac{m^2}{\mu^2}
	\right)\,m^2\hat{a}_1(x,x)
	-\ln\frac{m^2}{\mu^2}\,
	\hat{a}_2(x,x)\nonumber\\
	&&\qquad\qquad\qquad\qquad\qquad\left.
	+\sum_{n=1}^{\infty}\frac{(n-1)!}{m^{2n}}\,
	\hat{a}_{n+2}(x,x)\,\right\}.                     \label{4.2}
	\end{eqnarray}
Here ${\rm tr}$ denotes the trace over isotopic field indices, hats denote the
corresponding matrix structures in vector space of quantum fields and
$\hat{a}_n(x,x)$ are the Schwinger-DeWitt coefficients. The latter can be
systematically calculated for generic theory as spacetime invariants of
growing power in spacetime and fibre bundle curvatures
\cite{DW,Gilkey,BarvV,Avram}.

The situation with this expansion on a singular
instanton of Hawking and Turok is also not satisfactory -- all integrals of
curvature invariants starting with $\hat{a}_2(x,x)$ (quadratic in the
curvature and higher) diverge at the singularity. Fortunately, the lessons
from the asymptotic theory of semiclassical expansion teach us that the
lowest order terms can be trusted as long as they are well defined. In our
case this is the term quartic in masses with the logarithm. But this term
is exactly responsible for the dominant contribution to the anomalous
scaling (\ref{1.1}) quadratic in $|\xi|$. This term is dominating the quantum
part of effective action, while the others, although being divergent at the
singularity, are strongly suppressed by powers of $1/|\xi|$. With the
assumption that the back reaction of quantum stress tensor regulates these
divergences, one can conjecture that the quantum effective action on the
Hawking-Turok instanton in our model is still dominated by the anomalous
scaling term of eq.(\ref{1.1}). In the next section we analyze the
consequences of this conjecture.

\section{Energy scale, $N$ and $\Omega$ of open inflation}
\hspace{\parindent}
Consider the no-boundary and tunneling distribution functions (\ref{1.1})
with the Hawking-Turok action (\ref{3.16}) replacing $I(\varphi)$
and the anomalous scaling term.
A crucial difference from the case of closed cosmology is that the
$\varphi$-dependence of $I_{HT}(\varphi)$ is dominated now by
$|\xi|^0$-term which contains due to a big slowly varying logarithm
a large {\it positive} contribution quartic
in $m_P/\varphi$. Therefore, for large $|\xi|\gg 1$ the maximum of
this distribution exists only for the {\it no-boundary} state. The
corresponding peak is located at $\varphi=\varphi_I$, where $\varphi_I$
solves the equation
	\begin{eqnarray}
	\varphi_I^2\simeq 2\,\frac{m^2_P}{|\xi|}
	\,\left[\frac1{3\mbox{\boldmath$A$}}\ln\frac{2\pi|\xi|\varphi_I^2}
	{m_P^2}\right]^{1/2},\,\,
	|\delta|\ll 1.     \label{5.2}
	\end{eqnarray}
(To simplify equations we consider here and in what follows a small value
of $|\delta|$ and use approximate value of the numerical combination
$3/2-20/3\simeq-2\ln(18)$.) The solution of (\ref{5.2}) for small
$\mbox{\boldmath$A$}$ corresponds to the following parameters of the
probability peak -- mean value, the Hubble constant
$H(\varphi)\simeq\sqrt{\lambda/12|\xi|}\varphi$ and quantum dispersion
$\Delta\varphi\equiv[-d^2\ln\rho(\varphi_I)/d\varphi_I^2]^{-1/2}$
	\begin{eqnarray}
	&&\varphi^2_I\simeq 2\,\frac{m_P^2}{|\xi|}\,
	\left[\frac1{3\mbox{\boldmath$A$}}
	\ln\frac{8\pi}
	{\sqrt{54\mbox{\boldmath$A$}}}\right]^{1/2},\,\,\,
	H^2(\varphi_I)\simeq m^2_P\,\frac\lambda{|\xi|^2}\,
	\frac16\left[\frac1{3\mbox{\boldmath$A$}}
	\ln\frac{8\pi}
	{\sqrt{54\mbox{\boldmath$A$}}}\right]^{1/2},\nonumber\\
	&&\frac{\Delta\varphi}{\varphi_I}
	\sim\frac{\Delta H}{H}\sim
	\frac1{\sqrt{27\mbox{\boldmath$A$}}}
	\,\frac{\sqrt\lambda}{|\xi|}.                 \label{5.3}
	\end{eqnarray}
Similarly to the closed model, these parameters are suppressed relative to
the Planck scale by a small dimensionless ratio $\sqrt{\lambda}/|\xi|$ known
from the COBE normalization $\sqrt{\lambda}/|\xi|\sim \Delta T/T\sim 10^{-5}$.
However, in contrast with the closed model, this peak has a more
complicated dependence on the parameter $\mbox{\boldmath$A$}$.

To analyze the inflationary scenario generated by this peak we first use
the classical equations of motion. For the model (\ref{3.1}) they were
considered in much detail in \cite{efeq}. The slope of the
potential (\ref{3.8}) is positive for $\delta>-1$ which implies the
finite duration of the inflationary epoch with slowly decreasing inflaton
only in this range of $\delta$. The inflationary e-folding number in this case
is approximately given by the equation
	\begin{eqnarray}
	N\simeq\int_0^{\varphi_I}d\varphi\,
	\frac{3H^2(\varphi)}{|F(\varphi)|},             \label{5.4}
	\end{eqnarray}
where $F(\varphi)$ is the rolling force in the inflaton equation of motion
$\ddot\varphi+3H\dot\varphi-F(\varphi)=0$, $F(\varphi)=(2VU'-UV')/(U+3U'^2)
\sim -\lambda m_P^2(1+\delta)\varphi/48\pi\xi^2$. The
integration in (\ref{5.4}) gives
	\begin{eqnarray}
	N\simeq 12\pi\,\left[\frac1{3\mbox{\boldmath$A$}}
	\ln\frac{8\pi}
	{\sqrt{54\mbox{\boldmath$A$}}}\right]^{1/2}.   \label{5.5}
	\end{eqnarray}
Comparison of this result with the e-folding number, $N\sim 60$, necessary
for generating the observable density $\Omega$, $0<\Omega<1$, not very close
to one or zero, immeadiately gives the bound on $\mbox{\boldmath$A$}$
	\begin{eqnarray}
	\mbox{\boldmath$A$}\sim\frac{48\pi^2}{N^2}
	\ln\frac{2N}9\sim 0.3.                         \label{5.6}
	\end{eqnarray}
This bound justifies the validity of the slow roll approximation -- the
expansion in powers of $m_P^2/4\pi|\xi|\varphi^2\sim \ln N/N$ (see
eqs.(\ref{3.8}) and (\ref{3.13})).

These conclusions are based on classical equations of inflationary dynamics.
The latter should certainly be replaced by effective equations for mean fields
to have reliable answers within the same accuracy as the calculation of the
one-loop distribution function for the initial value of the inflaton. Since
quantum
effects qualitatively change the tree-level initial conditions, one should
expect that they might strongly influence the dynamics as well. Effective
equations of motion for the model (\ref{3.1}) were obtained in our recent
paper \cite{efeq}, but according to the discussion of the previous section
they are not, strictly speaking, applicable here. This is because the
Hawking-Turok instanton background does not satisfy the condition of the
local Schwinger-DeWitt expansion\footnote
{For closed cosmology with no-boundary or tunneling quantum states the
slow roll approximation guarantees the validity of the local Schwinger-DeWitt
expansion that was used in \cite{efeq} for the derivation of the effective
equations.
}.
It does not satisfy this condition globally, in the vicinity of the
singularity, but the open inflationary Universe lies inside the light cone
originating from the instanton pole antipoidal to the
singularity. Therefore, if we restrict ourselves with the local part
of effective equations, then for this part we can use our old
results \cite{efeq}. The influence of the spatially remote singularity
domain is mediated by nonlocal terms. These terms strongly depend on the
boundary conditions at the singular boundary and are beyond the control of
the local Schwinger-DeWitt approximation.
Within the same reservations as those concerning the finiteness of quantum
effects on this instanton (and the validity of the Hawking-Turok model as a
whole) we can neglect nonlocal terms and use only the local part of effective
equations.

There are two arguments in favour of this approximation. Firstly,
it is very likely that the nonlocal contribution of the singularity is
suppressed by inverse powers of $|\xi|\gg 1$. At least naively, these effects
are inverse to the size of the Universe given by the Hubble constant in
(\ref{5.3}) and also can be damped by $1/|\xi|$ in the same way as for the
singular part of the scalar curvature. Secondly, if we restrict
ourselves with a limited spatial domain of the very early open Universe
(close to the tip of the light cone originating from the regular pole of the
Hawking-Turok instanton), then these nonlocal terms do not contribute at all
in view of causality of effective equations, because at early moments of time
this domain is causally disconnected from singularity\footnote
{This argument is rigorous but somewhat macabre for the inhabitants of this
domain, because after a while they will suffer a fatal influence of fields
propagating from the singularity.
}.

Local quantum corrections in effective equations depend only on local geometry
of the quasi-DeSitter open Universe. They boil down to the replacement of
the classical coefficient functions ($V(\varphi),\,U(\varphi)$) of the model
(\ref{3.1}) by the effective ones calculated in \cite{efeq} for a wide class
of quantum fields coupled to the inflaton in the limit of big $|\xi|\gg 1$.
It remains to use these functions in classical equations and study the
inflation dynamics starting from the initial value of the inflaton
(\ref{5.2}). It turns out that unlike in the closed model (where the quantum
terms were of the same order of magnitude as classical ones) the quantum
corrections are strongly suppressed by the slow-roll parameter
$\sqrt{\mbox{\boldmath$A$}}/4\pi\sim \ln N/N$ already at the start of inflation.
In particular, the effective rolling force differs from the classical one
above by negligible correction
	\begin{eqnarray}
	F_{\rm eff}(\varphi)\simeq
	-\frac{\lambda m_P^2(1+\delta)}{48\pi\xi^2}\,\varphi\,
	\left\{1+\left[\frac{\mbox{\boldmath$A$}}{48\pi^2}
	\ln\frac{8\pi}{\sqrt{54\mbox{\boldmath$A$}}}
	\right]^{1/2}\frac{\varphi^2}{\varphi_I^2}\right\}.    \label{5.7}
	\end{eqnarray}
For comparison, in the closed model the second (quantum) term in curly
brackets enters with a unit coefficient of $\varphi^2/\varphi_I^2$, see
eqs.(6.7) and (6.10) of \cite{efeq}). This quantum correction gives
inessential contribution to the duration of the inflationary stage
(\ref{5.5}) and thus does not qualitatively change the above predictions
and bounds. The smallness of quantum corrections (roughly proportional to
$\mbox{\boldmath$A$}/32\pi^2$) can be explained by stronger bound on
$\mbox{\boldmath$A$}\sim 0.3$ (cf. $\mbox{\boldmath$A$}\leq 5.5$ in the
closed model \cite{efeq}) and another dependence of $\varphi_I$ on
$\mbox{\boldmath$A$}$.

Thus, the no-boundary quantum state on the Hawking-Turok instanton in the
model with large nonminimal curvature coupling generates open inflationary
scenario compatible with observations and, in particular, capable of
producing
the needed value of $\Omega$. No anthropic considerations or fine tuning of
initial conditions is necessary to reach such a final state of the Universe.
The only fine tuning we get is the bound on the parameter
$\mbox{\boldmath$A$}$ of the matter field sector (\ref{5.6}) and the
estimate of the ratio $\sqrt{\lambda}/|\xi|\sim 10^{-5}$ based on the
normalization from COBE which looks as a natural determination of coupling
constants of Nature from the experiment. The mechanism of such quantum birth
of the Universe is based on quantum effects on the Hawking-Turok instanton,
treated within semiclassical loop expansion. The validity of this expansion
is in its turn justified by the energy scale of the phenomenon (\ref{5.3})
which belongs to the GUT domain rather than the Planckian one.

\section*{Acknowledgements}
\hspace{\parindent}
The author is grateful to A.A.Starobinsky for
useful discussions. Helpful correspondence with A.Linde is also
deeply acknowledged. This work was supported
by the Russian Foundation for Basic Research under grants No 96-02-16287 and
No 96-02-16295, the European Community Grant INTAS-93-493-ext and
by the Russian Research program ``Cosmomicrophysics''.

\end{document}